\documentclass[prl,amsmath,showpacs,superscriptaddress]{revtex4}

\usepackage{amsmath}
\usepackage{amsbsy}
\usepackage{amscd}
\usepackage{amsopn}
\usepackage{amstext}
\usepackage{amsxtra}
\usepackage{graphicx}
\usepackage{epsfig}

\begin{document}
\title{High-accuracy approximation of binary-state dynamics   on networks}
\author{James P. Gleeson}
\affiliation{MACSI, Department of Mathematics \& Statistics, University of Limerick, Ireland.\\  james.gleeson@ul.ie}
\date{16 May 2011}

\begin{abstract}
Binary-state dynamics (such as the susceptible-infected-susceptible (SIS) model of disease spread, or Glauber spin dynamics)  on random networks are accurately approximated using master equations. Standard mean-field and pairwise theories are shown to result from seeking approximate solutions of the master equations. Applications to the calculation of SIS epidemic thresholds and critical points of non-equilibrium spin models are also demonstrated.
\end{abstract}
\maketitle

\begin{table}
\begin{center}
\begin{tabular}{|c|c|c|}
\hline
Process & $F_{k,m}$ & $R_{k,m}$\\
\hline
SIS \cite{SISrefs} & $\lambda m$ & $\mu$\\
Voter model \cite{Liggettbook} & $m/k$ &  $1- F_{k,m}$\\
Glauber dynamics \cite{Glauber63} & $\left[ 1+ \exp\left(\frac{2 J}{T}(k-2m)\right)\right]^{-1}$ & $1- F_{k,m}$\\
Majority-vote \cite{deOliveira92} & $\left\{ \begin{array}{cc}
Q & \text{ if }m<k/2 \\
1/2 & \text{ if }m=k/2\\
1-Q & \text{ if }m>k/2 \\
\end{array} \right. $&  $1- F_{k,m}$\\
 \hline
\end{tabular}
\end{center}
\caption{Infection and recovery rates for some examples of binary-state dynamics on networks: $k$ is the node's degree, $m$ is its number of infected neighbors. Parameters $\lambda$ and $\mu$ are SIS transmission and recovery rates; $T$ and $J$ are the temperature and interaction strength for the Ising model; $Q$ is the majority-vote noise parameter. Note $T=0$ Glauber dynamics are identical to those of the  $Q=0$  majority-vote model. }
\label{tabFR}
\end{table}

Dynamical processes running on complex networks are used to model a wide variety of phenomena \cite{Barratbook,Newmanbook}. Examples include spreading of diseases or opinions through a population  \cite{PastorSatorras01,Castellano09}, neural activity in the brain \cite{Honey09}, and cascading bank defaults in a financial system \cite{Haldane11}. The structure of the underlying network (e.g., its degree distribution) may strongly influence the dynamics and determine critical values of parameters (e.g., the critical temperature of the Ising spin model \cite{Dorogovtsev02}, or the epidemic threshold for disease-spread models \cite{Parshani10,Castellano10}). Accurate prediction of dynamics and critical points on networks of arbitrary degree distribution thus remains an important unsolved problem \cite{Barratbook}.

Mean-field theories (MF) are relatively simple to derive and can be quite accurate for dynamics on well-connected networks \cite{Dorogovtsev08}. However, on sparse networks, or close to critical points, MF theories perform poorly (see, for example, Fig.~1 below). Pair-wise approximations (PA), which take into account the states of both nodes at the ends of a network edge, improve on MF, but have been derived for fewer dynamical processes (examples are \cite{Eames02,Pugliese09}). In this Letter we demonstrate a tractable master equation approach for binary-state dynamics, with accuracy exceeding both MF and PA. We show that PA and MF theories may be derived by seeking approximate solutions of the master equations. We write down the explicit PA equations for the general case, thus giving the first derivation of pair-wise approximations for a range of dynamical processes. Finally, we use the master equations to calculate critical points such as the epidemic threshold for the susceptible-infected-susceptible (SIS) model (or contact process), and the critical noise level in the majority-vote model \cite{deOliveira92}.

We consider binary-state dynamics  on static, undirected, connected networks in the limit of infinite network size. For convenience, we call the two possible states of a node \emph{susceptible} and \emph{infected}, as is common in disease-spread models. However, this approach also applies to other binary-state dynamics, such as spin models \cite{deOliveira93},  where each node may be in the $+1$ (spin-up=infected) or the $-1$ (spin-down=susceptible) state.
 The networks have degree distribution $P_k$ and are generated by the configuration model \cite{Newmanbook}. Dynamics are stochastic, and are defined by infection and recovery probabilities which depend on the degree $k$ of a node, and on the current number $m$ of infected neighbors of the node. Thus $F_{k,m}\,dt$ is defined as the probability that a $k$-degree node that is susceptible at time $t$, with $m$ infected neighbors, changes its state to infected by time $t+dt$, where $dt$ in an infinitesimally small time interval. Similarly, $R_{k,m}\,dt$ is  the probability that a $k$-degree infected node with $m$ infected neighbor moves to the susceptible state within a time $dt$. These general infection and recovery probabilities can describe many dynamical processes of interest, see Table~\ref{tabFR} for some examples.

Approximate master equations for dynamics of this type can be derived by generalizing the approach used in \cite{Marceau10} for SIS dynamics, see  Appendix A.
Let  $s_{k,m}(t)$ (resp. $i_{k,m}(t)$) be the fraction of $k$-degree nodes that are susceptible (resp. infected) at time $t$, and have $m$ infected neighbors. Then the fraction $\rho_k(t)$ of $k$-degree nodes that are infected at time $t$ is given by
$
\rho_k(t) = \sum_{m=0}^k i_{k,m} = 1 - \sum_{m=0}^k s_{k,m}, 
$
and the fraction of infected nodes in the whole network is found by summing over all $k$-classes:
$
\rho(t) = \left<\rho_k(t)\right>\equiv \sum_k P_k \, \rho_k(t).
$

The master equations for the evolution of $s_{k,m}(t)$ and  $i_{k,m}(t)$ are (see Appendix A):
\begin{widetext}
\begin{eqnarray}
\frac{d}{d t}s_{k,m} &=& -F_{k,m} s_{k,m} + R_{k,m} i_{k,m} - \beta^s (k-m) s_{k,m} + \beta^s(k-m+1) s_{k,m-1} - \gamma^s m s_{k,m} + \gamma^s(m+1) s_{k,m+1}, \label{seqns}\\
\frac{d}{d t}i_{k,m} &=& -R_{k,m} i_{k,m} + F_{k,m} s_{k,m} - \beta^i (k-m) i_{k,m} + \beta^i(k-m+1) i_{k,m-1} - \gamma^i m i_{k,m} + \gamma^i(m+1) i_{k,m+1}, \label{ieqns}
\end{eqnarray}
\end{widetext}
for each $m$ in the range $0,\ldots,k$, and for each $k$-class in the network.
 The first two terms on the right hand side of each equation represent transitions due to infection or recovery of a $k$-degree node. The remaining four terms account for infection or recovery of a neighbor. The rates $\beta^s$, $\gamma^s$, $\beta^i$, and $\gamma^i$ are approximated by tracking the number of edges of each type.
 To calculate $\beta^s$, for example, we count the number of $S$-$S$ edges (i.e., edges between two susceptible nodes) in the network at time $t$, and then count the number of edges which switch from being $S$-$S$ edges to $S$-$I$ edges in the time interval $dt$; the probability $\beta^s\, dt$ is given by taking the ratio of the latter to the former, giving $\beta^s = \frac{\left<\sum_{m=0}^k (k-m) F_{k,m} \,s_{k,m}\right>}{\left< \sum_{m=0}^k (k-m) s_{k,m}\right>} $. Similarly, we have
$\gamma^s =  \frac{\left<\sum_{m=0}^k (k-m) R_{k,m} \,i_{k,m}\right>}{\left< \sum_{m=0}^k (k-m) i_{k,m}\right>}  $, $\beta^i = \frac{\left< \sum_{m=0}^k m\, F_{k,m} \,s_{k,m}\right>}{\left< \sum_{m=0}^k m\, s_{k,m}\right>}$, and $\gamma^i =  \frac{\left< \sum_{m=0}^k m\, R_{k,m} \,i_{k,m}\right>}{\left< \sum_{m=0}^k m\, i_{k,m}\right>}$, see Appendix A for details.

\begin{figure}[b]
\centering
\epsfig{figure=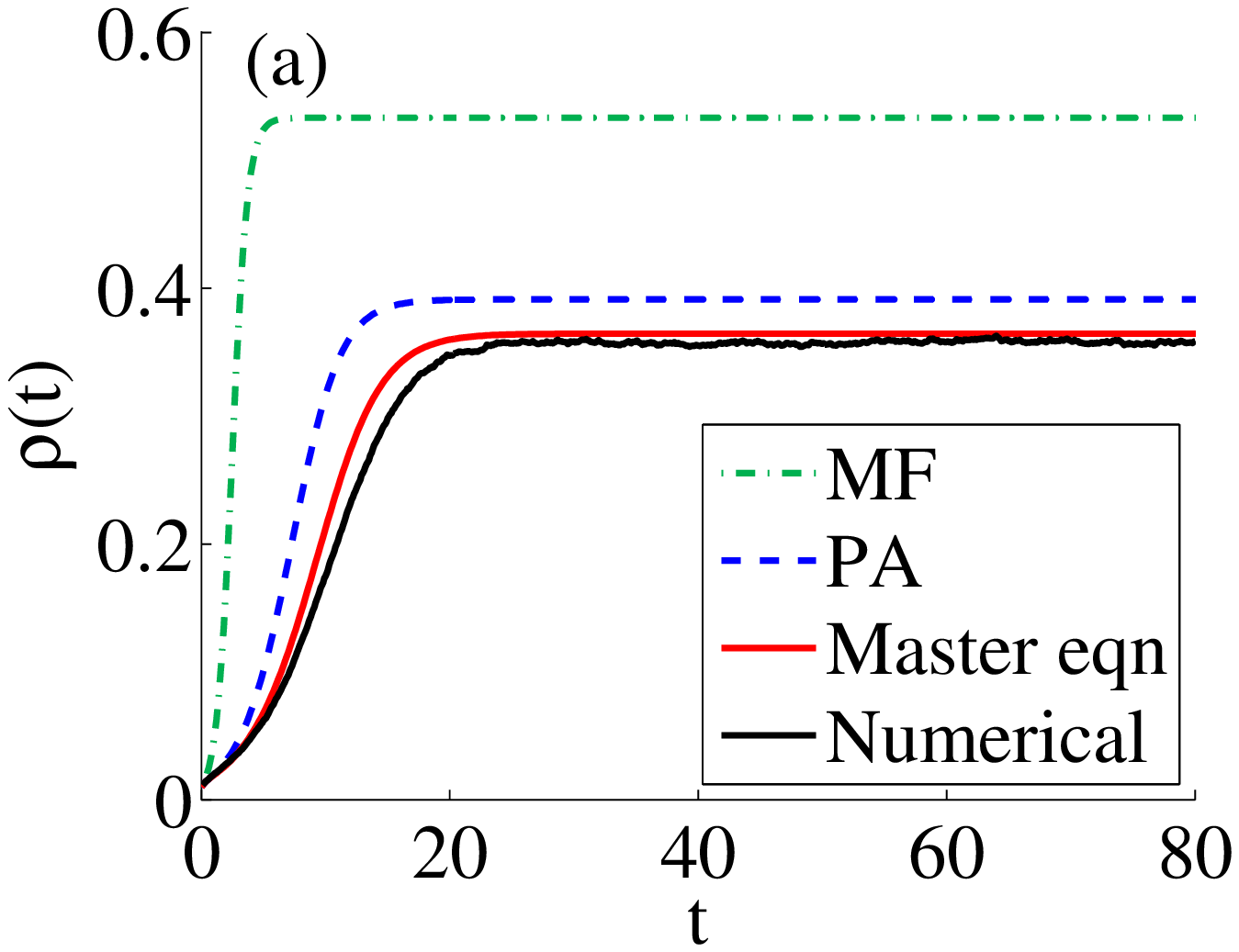,width=4.25cm}
\epsfig{figure=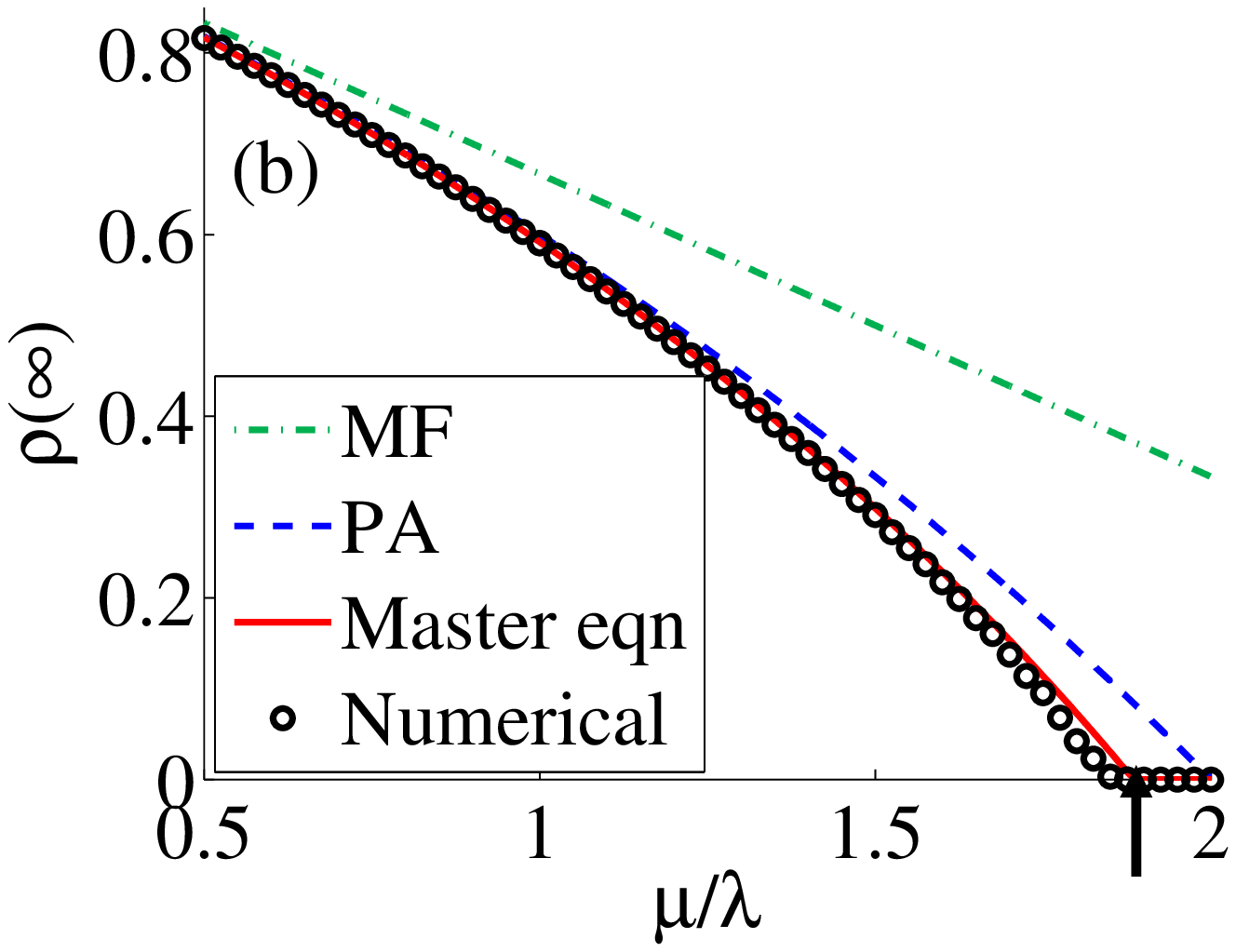,width=4.25cm}
\caption{ (a) Infected fraction $\rho(t)$ in the SIS disease spread model on 3-regular random graphs, with transmission rate $\lambda=1$ and recovery rate $\mu=1.4$. (b) Steady-state fraction of infected nodes as a function of the non-dimensional recovery rate $\mu/\lambda$. The arrow marks the epidemic threshold predicted from the linearized master equations (top row of Table~\ref{tab1}(a)).
}\label{fig4}
\end{figure}

The master equations (\ref{seqns}) and (\ref{ieqns}), with the time-dependent rates $\beta^s$, $\gamma^s$, $\beta^i$ and $\gamma^i$ (defined as nonlinear functions of $s_{k,m}$ and $i_{k,m}$), form a closed system of deterministic equations which can be solved numerically using standard methods. Assuming a randomly-chosen fraction $\rho(0)$ of nodes are initially infected,  the initial conditions are
$
s_{k,m}(0)=\left( 1- \rho(0) \right) B_{k,m}(\rho(0))$, $i_{k,m}(0) = \rho(0) B_{k,m}(\rho(0)), 
$
where $B_{k,m}(q)$ denotes the binomial factor
$\binom{k}{m} q^m (1-q)^{k-m}$.
Note that the evolution equations are completely prescribed by the functions $F_{k,m}$ and $R_{k,m}$, and so this method can be applied to any stochastic dynamical process defined by transition rates of this type. For the SIS model, equations (\ref{seqns}) and (\ref{ieqns}) were derived in \cite{Marceau10}, but were not analyzed as here.

\begin{figure}[b] 
\centering
\epsfig{figure=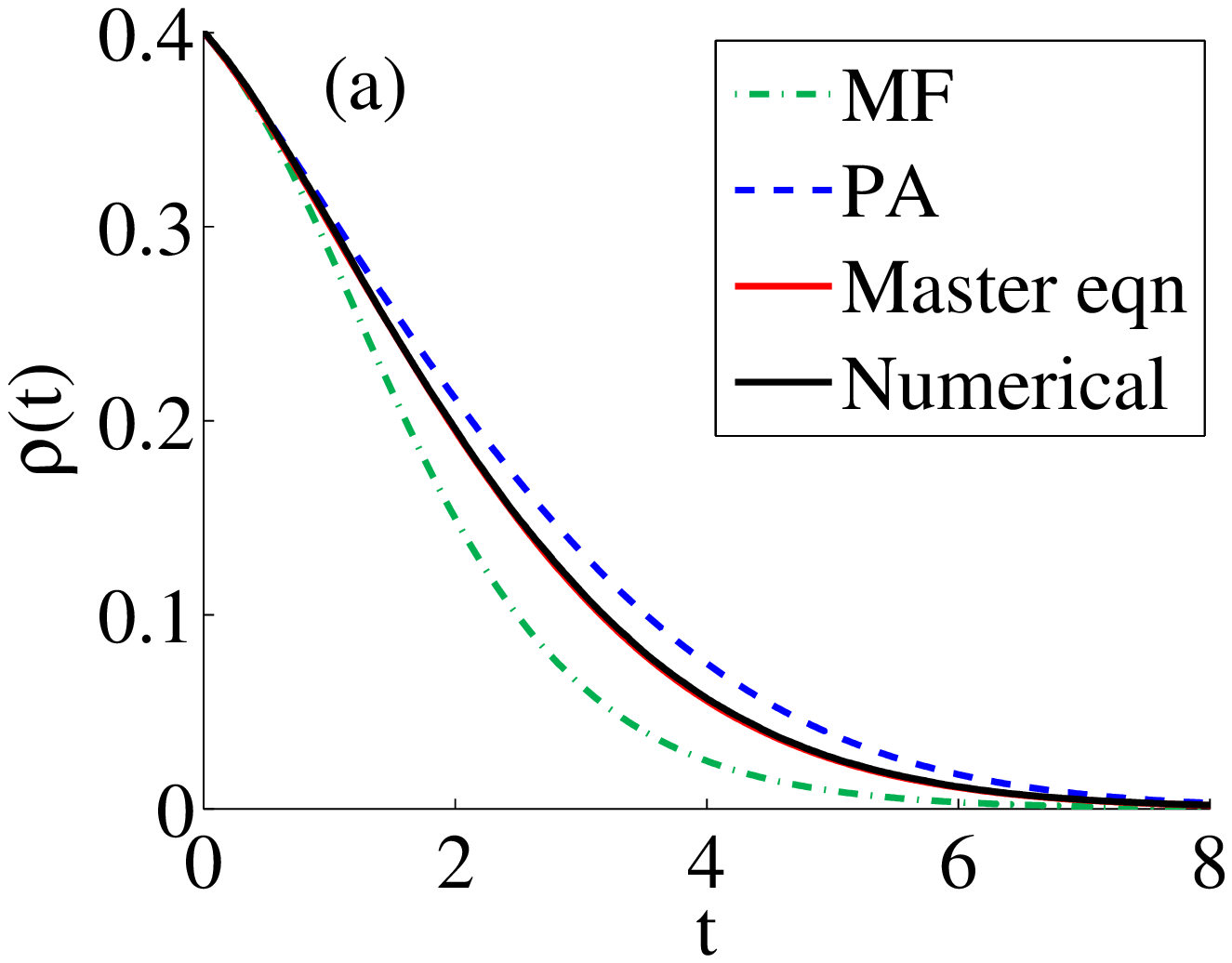,width=4.25cm}
\epsfig{figure=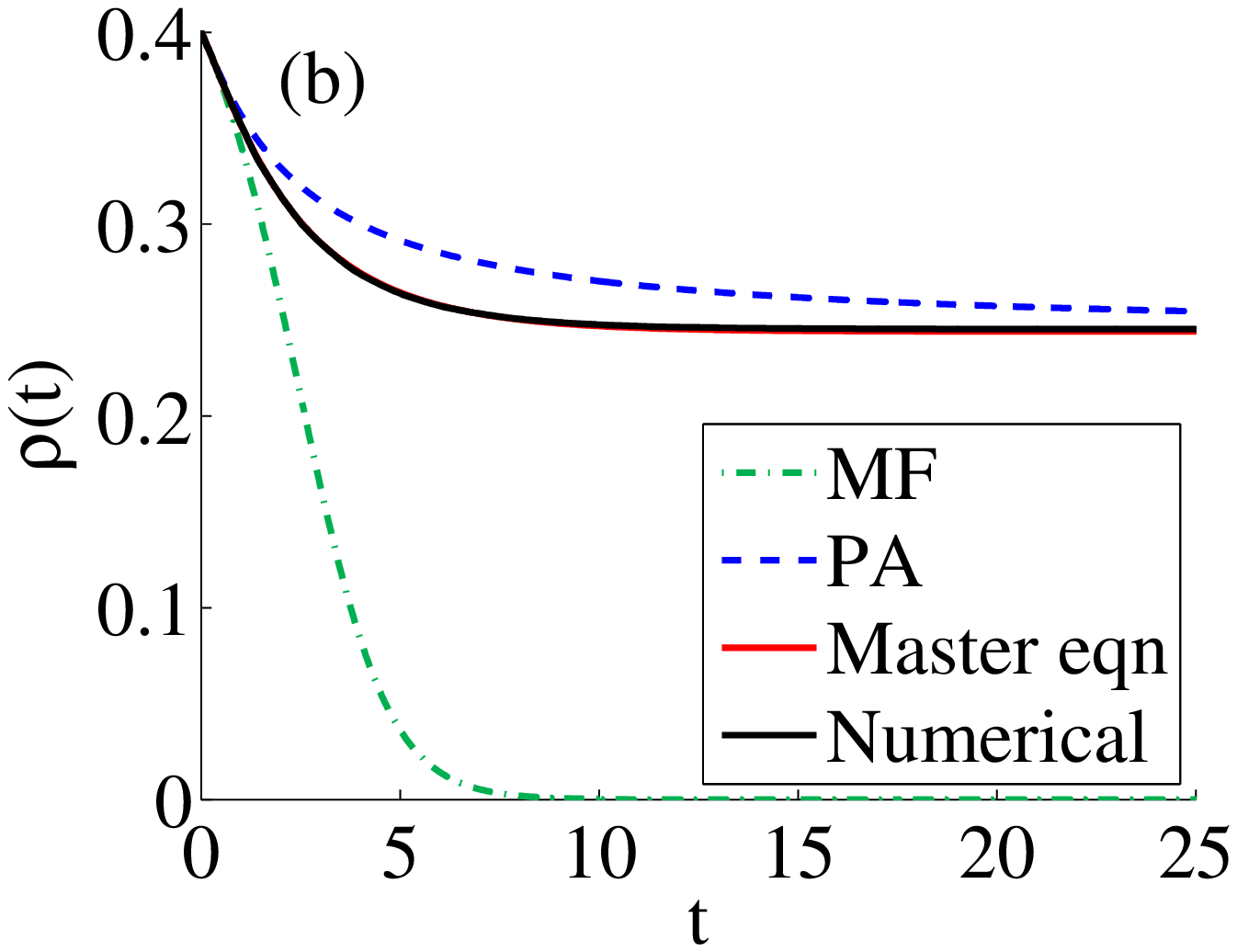,width=4.25cm}
\caption{Infected fraction $\rho(t)$ (i.e., fraction of $+1$ spins) for zero-temperature Glauber dynamics on (a) networks with truncated power-law degree distribution: $P_k \propto k^{-2.5}$ for $3 \le k \le 20$, and (b) 3-regular random graphs. In each case the initial condition is $\rho(0)=0.4$.
}\label{fig3}
\end{figure}

Figure 1(a) shows the infected fraction $\rho(t)$ of nodes in the SIS model run on a 3-regular random graph (i.e., a Bethe lattice, with $P_k=\delta_{k,3}$). The master equations (\ref{seqns})--(\ref{ieqns}) clearly give a better approximation to the actual stochastic dynamics than standard methods (here, the mean-field theory of \cite{PastorSatorras01} and the pair-approximation method of \cite{Levin96, Eames02}---note these are reproduced by equations (\ref{MF}) and (\ref{PA}) below).
The steady-state infected fraction is plotted as a function of the non-dimensional recovery rate $\mu/\lambda$ in Fig.~1(b). The master equation solutions give a significantly better estimate of the epidemic threshold than the standard approximations: we pursue this further below. Figures \ref{fig3}(a) and \ref{fig3}(b) demonstrate that similar conclusions hold for zero-temperature Glauber dynamics \cite{Castellano05} on networks with truncated power-law degree distributions and on 3-regular random graphs. Here the comparison is with the mean-field theory of \cite{Castellano06} (see also (\ref{MF}) below), and the pair approximation from equation (\ref{PA}) below.
Figure 2(b) shows that our approach captures the fact that $T=0$ Glauber dynamics on networks can freeze in disordered states; this phenomenon is not captured at all by MF \cite{Castellano06}.

 For dynamics on a general network, with non-empty degree classes from $k=0$ up to a cutoff $k_\text{max}$, the number of differential equations in the system (\ref{seqns})--(\ref{ieqns}) is $(k_\text{max}+1)(k_{\text{max}}+2)$, and so grows with the square of the largest degree.
In certain no-recovery cases (i.e., $R_{k,m}\equiv 0$), such as Watts' threshold model \cite{Watts02}, $k$-core size calculations \cite{Dorogovtsev06}, and bootstrap percolation \cite{Baxter10}, we can show that an \emph{exact} solution of the master equations is obtained by solving just two differential equations (as given in \cite{Gleeson08a}). For general dynamics, however,  some approximation is necessary if it is desirable to reduce the master equations to a lower-dimensional  system. One possibility is to consider the parameters $p_k(t)$ (resp. $q_k(t)$), defined as the probability that  a randomly-chosen neighbor of a susceptible (resp. infected) $k$-degree node is infected at time $t$. Noting that $p_k(t)$ can be expressed in terms of $s_{k,m}$ as $\sum_{m=0}^k m s_{k,m}/\sum_{m=0}^k k s_{k,m}$,
an evolution equation for $p_k$ may be derived by multiplying equation (\ref{seqns}) by $m$ and summing over $m$. The right-hand-side of the resulting equation contains higher moments of $s_{k,m}$, so a closure approximation is needed to proceed. If we make the ansatz that $s_{k,m}$ and $i_{k,m}$ are proportional to binomial distributions:
$
s_{k,m}\approx (1-\rho_k)\,B_{k,m}(p_k)$, $i_{k,m} \approx \rho_k\, B_{k,m}(q_k), 
$, we obtain the \emph{pair approximation (PA)}, consisting of the $3k_\text{max}+1$ differential equations:
\begin{widetext}
\begin{eqnarray}
\frac{d}{d t}\rho_k &=& -\rho_k \sum_{m=0}^k R_{k,m} B_{k,m}(q_k) + (1-\rho_k)\sum_{m=0}^k F_{k,m}B_{k,m}(p_k),\nonumber\\
\frac{d}{d t}p_k &=& \sum_{m=0}^k \left[ p_k - \frac{m}{k}\right]\left[ F_{k,m} B_{k,m}(p_k) - \frac{\rho_k}{1-\rho_k} R_{k,m} B_{k,m}(q_k)\right] + \overline{\beta^s}(1-p_k)-\overline{\gamma^s}p_k,\nonumber\\
\frac{d}{d t}q_k &=& \sum_{m=0}^k \left[ q_k - \frac{m}{k}\right]\left[ R_{k,m} B_{k,m}(q_k) - \frac{1-\rho_k}{\rho_k} F_{k,m} B_{k,m}(p_k)\right] + \overline{\beta^i}(1-q_k)-\overline{\gamma^i}q_k , \label{PA}
\end{eqnarray}
\end{widetext}
for each $k$-class. The rates here are given by  inserting the binomial ansatz into the general formulas, so that $\overline{\beta^s}$, for example, is 
$\left<(1-\rho_k)\sum_m (k-m) F_{k,m}B_{k,m}(p_k)\right>/\left<(1-\rho_k)k(1-p_k)\right>$;
 initial conditions are $\rho_k(0)=p_k(0)=q_k(0)=\rho(0)$.

 A cruder, \emph{mean-field (MF)}, approximation results from replacing both $p_k$ and $q_k$ with $\omega$:
$
s_{k,m}\approx (1-\rho_k)\,B_{k,m}(\omega)$, $i_{k,m} \approx \rho_k\, B_{k,m}(\omega), 
$
where $\omega=\left< \frac{k}{z} \rho_k\right>$ is the probability that one end of a randomly-chosen edge is infected.
Using this ansatz in the master equations yields a closed system of $k_\text{max}+1$ differential equations for the fraction $\rho_k$ of infected $k$-degree nodes:
\begin{equation}
\frac{d}{d t}\rho_k= -\rho_k \sum_{m=0}^k R_{k,m} B_{k,m}(\omega) + (1-\rho_k)\sum_{m=0}^k F_{k,m}B_{k,m}(\omega), \label{MF}
\end{equation}
with $\rho_k(0)=\rho(0)$.

 The PA and MF approximations (\ref{PA}) and (\ref{MF}) yield increasingly simpler systems of equations for any process that can be expressed in terms of infection and recovery rates $F_{k,m}$ and $R_{k,m}$. For the SIS model, the PA equations (\ref{PA}) are  those of Eames and  and Keeling \cite{Eames02}, while the MF equations (\ref{MF}) are precisely those of  Pastor-Satorras and Vespignani \cite{PastorSatorras01}. For the voter model \cite{Liggettbook}, the MF equations (\ref{MF})  reduce to those in \cite{Sood05},
while the PA equations (\ref{PA}) lie between those of \cite{Pugliese09} and \cite{Vazquez08} in terms of complexity.
 The MF equations (\ref{MF}) for zero-temperature Glauber dynamics reproduce the mean-field theory of \cite{Castellano06} (in the limit of infinite network size). For this and related non-equilibrium spin models, such as the majority-vote model, steady-state PA equations for the special case of 4-regular graphs (i.e., $P_k=\delta_{k,4}$) are derived in \cite{deOliveira93}. However, to our knowledge, no PA equations such as (\ref{PA}) have been derived for these dynamics on networks with arbitrary degree distribution $P_k$.
 Note also that a coarser type of PA, using the ansatz $s_{k,m}=(1-\rho_k)B_{k,m}(p)$, $i_{k,m}=\rho_k B_{k,m}(q)$ (i.e., with $k$-independent parameters $p$ and $q$) gives  the equations recently derived in \cite{House11} for SIS, and those in \cite{Vazquez08} for the voter model. See Appendix B for details of this ``homogeneous'' PA.

 We briefly highlight another important application of the master equations: the calculation of the epidemic threshold for the SIS disease spread model \cite{Parshani10,Castellano10}. If the seed fraction of infected nodes $\rho(0)$ is sufficiently small, an appropriate linearization of the master equations (\ref{seqns})--(\ref{ieqns}) determines whether the infected fraction will grow (to epidemic proportions), or will decay to zero. This reduces the problem to linear stability analysis, and so to the calculation of the largest eigenvalue of a matrix (with dimension of order $k_\text{max}^2$). In Table~\ref{tab1}(a) we show the critical values of the parameter $\mu/\lambda$ for SIS dynamics on $z$-regular random graphs calculated in this way, and compare with the explicit values predicted by PA \cite{Levin96,Eames02} and MF \cite{PastorSatorras01} methods (i.e., $z-1$ and $z$, respectively). Recently it was argued that SIS infection can persist indefinitely in networks containing nodes of sufficiently high degree, due to recurring reinfections between hub nodes and their neighbors \cite{Castellano10}. The master equation formalism does not capture this effect, because the definitions of the rates ($\beta^s$, $\gamma^s$, etc) use global counts of edge types, and so wash out structural correlations specific to the immediate neighborhood of hub nodes.

Linear stability analysis  may also be applied to spin models with up-down symmetry, which have the property
$
R_{k,m} = 1- F_{k,m}= F_{k,k-m},
$, and where the magnetization $M(t)$ (the average of all spins in the network) is given  by $M=2\rho-1$. Stability analysis of the (disordered) fixed point with $\rho=1/2$ gives the location of critical points marking the transition between disordered and ordered phases.  Applying this method to Glauber dynamics reproduces the results of \cite{Dorogovtsev02} for the critical temperature of the Ising model. It also accurately approximates numerically-determined critical values for non-equilibrium spin models, such as the critical noise $Q_c$ in the majority-vote model, see Table II(b).

\begin{table}
\begin{center}
\begin{tabular}{|c|c|c|c|c||c|c|c|c|c|}
\hline
\multicolumn{5}{|c||}{(a) SIS, $z$-regular}&
\multicolumn{5}{|c|}{(b) Majority-vote, PRG}\\
\hline
$z$ & bounds &Master & PA & MF & $z$ & num  & Master & PA & MF\\
    &  \cite{Pemantle92} &  eqn   & \cite{Eames02}   & \cite{PastorSatorras01}        & & \cite{Pereira05}     &  eqn  &    &    \\ 
\hline
 3 & $(1.65,\,2)$  & 1.88& 2& 3& 3 & 0.135& 0.137 & 0.141 & 0.180   \\
 4 &$(2.56,\,3)$& 2.91 & 3 & 4  & 4 & 0.181& 0.184 & 0.185 & 0.214  \\
 5 &$(3.58,\,4)$& 3.93 & 4 & 5 & 6 & 0.240& 0.242 & 0.242 & 0.259\\
 10&$(8.63,\,9)$ & 8.97 & 9 & 10& 8& 0.275& 0.277 & 0.276& 0.288\\
 \hline
\end{tabular}
\end{center}
\caption{(a) Critical values of $\mu/\lambda$ for epidemic spread in the SIS model on $z$-regular graphs.
  Lower and upper bounds for the critical value of $\mu/\lambda$ for the contact process on a tree (defined as the largest value of $\mu/\lambda$ for which the infection survives forever with positive probability) are from \cite{Pemantle92}. Note that the largest eigenvalue of the adjacency matrix for these networks is $\lambda_1=z$, so the method of Prakash \emph{et al.} \cite{Prakash10} gives the same (inaccurate) prediction for the critical value as MF theory.
  (b) Critical value of the noise parameter $Q$ in the majority-vote model on Poisson (Erd\"{o}s-R\'{e}nyi) random graphs of mean degree $\left< k \right>=z$. Numerical values are from \cite{Pereira05}, other values are determined via stability analysis of equations (\ref{seqns})--(\ref{MF}).
}
\label{tab1}
\end{table}

In summary, we have derived the master equations (\ref{seqns})--(\ref{ieqns})---first introduced for SIS dynamics in \cite{Marceau10}---for general binary-state dynamics on networks, and demonstrated that their accuracy supersedes standard MF and PA methods. Mean-field and pairwise theories are derived as approximate solutions of the master equations, and equations (\ref{PA}) explicitly give pair approximations for any dynamics defined by infection and recovery  rates  $F_{k,m}$ and $R_{k,m}$. Finally, we demonstrated the application of the master equations to calculating epidemic thresholds and critical parameter values via linear stability analysis, improving significantly on existing MF and PA estimates.

  We anticipate further applications of the master equation approach to the calculation of critical points in opinion models and spin systems, and expect possible extensions to include multiple-state dynamics (such as the SIR disease-spread model \cite{Noel09,Ball08,  Marceau11, TomeZiff10}), multiple node types \cite{Goltsev10}, discrete-time dynamics \cite{Gomez10}, and network models with non-zero clustering  \cite{Newman09, Miller09, Gleeson09,HebertDufresne10, Hackett10}.

This work was funded by Science Foundation Ireland awards 06/IN.1/I366 and MACSI
06/MI/005. Helpful discussions with Sergey Melnik, Rick Durrett, and Claudio Castellano, and participants at the SAMSI Dynamics on Networks workshop are gratefully acknowledged.

\appendix
\section{Appendix A: Derivation of Master equations}
[The material in Appendices A and B appeared in the early ArXiv versions of this paper.]
\begin{figure*}
\centering
\epsfig{figure=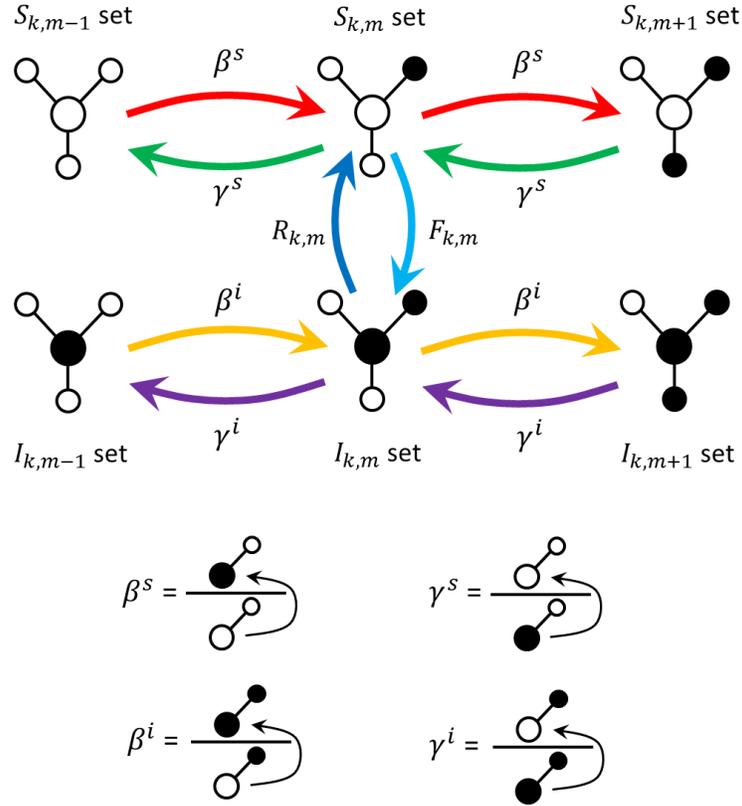,width=10cm}
\caption{Schematic of transitions to/from the $S_{k,m}$ and $I_{k,m}$ sets, as described in equations (\ref{Weqn}) through (\ref{betai}). For each set, the central (Ego) node is shown along with some of its neighbors: black nodes are infected, white nodes are susceptible. See also Fig.~1 of \cite{Marceau10}.}\label{fig_schematic}
\end{figure*}

We consider binary-state dynamics  on static, undirected, connected networks in the limit of infinite network size (i.e., $N\to \infty$, where $N$ is the number of nodes in the network). For convenience, we call the two possible node states \emph{susceptible} and \emph{infected}, as is common in disease-spread models. However, our approach also applies to other binary-state dynamics, such as spin systems \cite{deOliveira93},  where each node may be in the $+1$ (spin-up) or the $-1$ (spin-down) state.
 The networks have degree distribution $P_k$ and are generated by the configuration model \cite{Newmanbook,Bender78, Bollobas80}. Dynamics are stochastic, and are defined by infection and recovery probabilities which depend on the degree $k$ of a node, and on the current number $m$ of infected neighbors of the node. Thus $F_{k,m}\,dt$ is defined as the probability that a $k$-degree node that is susceptible at time $t$, with $m$ infected neighbors, changes its state to infected by time $t+dt$, where $dt$ in an infinitesimally small time interval. Similarly, $R_{k,m}\,dt$ is  the probability that a $k$-degree infected node with $m$ infected neighbor moves to the susceptible state within a time $dt$. These general infection and recovery probabilities can describe many dynamical processes on networks. For example, in the susceptible-infected-susceptible (SIS) model of disease spread (the contact process) \cite{SISrefs}, each susceptible node may be infected at a rate $\lambda$ by each infected neighbor, and each infected node recovers at a constant rate $\mu$,  so the rates $F_{k,m}$ and $R_{k,m}$ take the form
 \begin{equation}
F^{\text{SIS}}_{k,m}=\lambda\, m, \quad \quad  R^{\text{SIS}}_{k,m}=\mu.
 \end{equation}
 Zero-temperature Glauber dynamics for a spin system on a network \cite{Glauber63,Castellano05, Castellano06} provide another example.
 Here, each node has a spin of $+1$ or $-1$  (which we can identify with the infected or susceptible state, respectively). In each infinitesimal time step, one node is selected at random ($dt=1/N$) and its spin is set to $+1$ if the local field (sum of its neighbors' spins) is positive, to $-1$ is the local field is negative, and to $\pm 1$ with equal probability if the local field is zero. Thus,
 the new spin matches the majority of its neighbors' spins (with a random choice in case of a tie). The infection and recovery rates may therefore be expressed as
\begin{equation}
F^\text{Glauber}_{k,m} = \left\{ \begin{array}{cl}
                0 & \text{ if }m<k/2,\\
                \frac{1}{2} & \text{ if }m=k/2,\\
                1 & \text{ if }m>k/2,
                \end{array}\right.
 \quad\quad
  R^\text{Glauber}_{k,m} = \left\{ \begin{array}{cl}
                1 & \text{ if }m<k/2,\\
                \frac{1}{2} & \text{ if }m=k/2,\\
                0 & \text{ if }m>k/2.
                \end{array}\right.
\end{equation}
The voter model \cite{Liggettbook} has rates given by
\begin{equation}
F_{k,m}^{\text{voter}} = \frac{m}{k}, \quad\quad R_{k,m}^\text{voter} = \frac{k-m}{k},
\end{equation}
since infection of a $k$-degree node, for example, occurs by copying one of $m$ infected neighbors out of $k$ possible choices.

We now proceed to derive the master equations for dynamics of this type, closely following the approach used in \cite{Marceau10} for SIS dynamics.
Let $S_{k,m}$ (resp. $I_{k,m}$) be the set of nodes which are susceptible (resp. infected), have degree $k$, and have $m$ infected neighbors. To quantify the size of these sets, define $s_{k,m}(t)$ (resp. $i_{k,m}(t)$) as the fraction of $k$-degree nodes that are susceptible (resp. infected) at time $t$, and have $m$ infected neighbors. Then the fraction $\rho_k(t)$ of $k$-degree nodes that are infected at time $t$ is given by
\begin{equation}
\rho_k(t) = \sum_{m=0}^k i_{k,m} = 1 - \sum_{m=0}^k s_{k,m}, \label{rhok}
\end{equation}
and the fraction of infected nodes in the whole network is found by summing over all $k$-classes:
\begin{equation}
\rho(t) = \sum_k P_k \, \rho_k(t).
\end{equation}
If a randomly-chosen fraction $\rho(0)$ of nodes are initially infected, then the initial conditions for $s_{k,m}$ and $i_{k,m}$ are easily seen to be
\begin{equation}
s_{k,m}(0)=\left( 1- \rho(0) \right) B_{k,m}(\rho(0)), \quad\quad i_{k,m}(0) = \rho(0) B_{k,m}(\rho(0)), \label{ICs}
\end{equation}
where we introduce the convenient notation $B_{k,m}(q)$ for the binomial factor $\left(\!\!\begin{array}{c} k\\m \end{array} \!\!\right) q^m (1-q)^{k-m}$. Note that we can also calculate the number of edges of various types using this formalism. For example, the number of edges in the network which join a susceptible node to an infected node (we call these \emph{$S$-$I$ edges} for short) can be expressed in two equivalent ways:
\begin{equation}
N \sum_k P_k \sum_{m=0}^k m\, s_{k,m}\quad \text{ or } \quad N \sum_k P_k \sum_{m=0}^k (k-m) \,i_{k,m}. \label{star}
\end{equation}
The first of these expressions, for example, follows from noting that in a sufficiently large network that there are $N P_k$ nodes of degree $k$, of which a fraction $s_{k,m}$ are susceptible and have $m$ infected neighbors. Each such node contributes $m$ edges to the total number of $S$-$I$ edges. Similar expressions may also be given for the number of $S$-$S$ and $I$-$I$ edges in the network. We note that the equivalence of the two expressions in (\ref{star}) is preserved by the evolution equations described below.

Next, we examine how the size of the $S_{k,m}$ set changes in time. We write the general expression
\begin{eqnarray}
s_{k,m}(t+dt) &=& s_{k,m}(t) - W(S_{k,m}\!\to\! I_{k,m})\,s_{k,m}\,dt+W(I_{k,m}\!\to\! S_{k,m}) \, i_{k,m}\, dt \nonumber\\
& & \hspace{1cm}- W(S_{k,m}\!\to\! S_{k,m+1}) \, s_{k,m}\, dt + W(S_{k,m-1}\!\to\! S_{k,m})\, s_{k,m-1}\, dt \nonumber\\
& & \hspace{1cm}- W(S_{k,m}\!\to\! S_{k,m-1}) \, s_{k,m}\, dt + W(S_{k,m+1}\!\to\! S_{k,m})\, s_{k,m+1}\, dt \label{Weqn}
\end{eqnarray}
to reflect all the transitions whose rate is linear in $dt$ (all other state-transitions are negligible in the $dt \to 0$ limit), see Fig.~\ref{fig_schematic}. Here $W(S_{k,m}\to I_{k,m})\, dt$, for example, is the probability that a node in the $S_{k,m}$ set at time $t$ moves to the $I_{k,m}$ set by time $t+dt$. It is clear from the definitions above that
\begin{equation}
W(S_{k,m}\!\to\!I_{k,m}) = F_{k,m} \quad \text{ and }\quad W(I_{k,m}\!\to\!S_{k,m}) = R_{k,m}.
\end{equation}
A node moves from the $S_{k,m-1}$ set to the $S_{k,m}$ set if it remains susceptible, while one of its susceptible neighbors becomes infected. Note this means that an $S$-$S$ edge changes to an $S$-$I$ edge. If we suppose that $S$-$S$ edges change to $S$-$I$ edges at a (time-dependent) rate $\beta^s$, we can write \footnote{The main approximation here is to assume that the edge-state transition rate $\beta^s$ is the same for all $S$-$S$ edges in the network, regardless of their local neighborhood---the same assumption is made for the other rates $\gamma^s$, $\beta^i$, and $\gamma^i$. See also the explanation in \cite{Marceau10} for the SIS case.}
\begin{equation}
W(S_{k,m}\!\to\!S_{k,m+1})=\beta^s(k-m)\quad \text{ and }\quad W(S_{k,m-1}\!\to\!S_{k,m}) = \beta^s (k-m+1),
\end{equation}
since nodes in the $S_{k,m}$ set have $k-m$ susceptible neighbors, while those in the $S_{k,m-1}$ set have $k-m+1$ susceptible neighbors. To calculate $\beta^s$, we count the number of $S$-$S$ edges in the network at time $t$, and then count the number of edges which switch from being $S$-$S$ edges to $S$-$I$ edges in the time interval $dt$; the probability $\beta^s\, dt$ is given by taking the ratio of the latter to the former, i.e.
\begin{equation}
\beta^s\, dt = \frac{\sum_k P_k \sum_{m=0}^k (k-m) F_{k,m} \,s_{k,m}\, dt}{\sum_k P_k \sum_{m=0}^k (k-m) s_{k,m}} . \label{betas}
\end{equation}
A similar approximation is used to define $\gamma^s$, the rate at which $S$-$I$ edges change to $S$-$S$ edges due to the recovery of an infected node:
\begin{equation}
\gamma^s =  \frac{\sum_k P_k \sum_{m=0}^k (k-m) R_{k,m} \,i_{k,m}}{\sum_k P_k \sum_{m=0}^k (k-m) i_{k,m}} , \label{gammas}
\end{equation}
and we then write
\begin{equation}
W(S_{k,m}\!\to\!S_{k,m-1})=\gamma^s m\quad \text{ and }\quad W(S_{k,m+1}\!\to\!S_{k,m}) = \gamma^s (m+1).
\end{equation}
Taking the limit $dt\to 0$ of equation (\ref{Weqn}) gives the master equation for the evolution of $s_{k,m}(t)$ (see Fig.~\ref{fig_schematic}):
\begin{equation}
\frac{d}{d t}s_{k,m} = -F_{k,m} s_{k,m} + R_{k,m} i_{k,m} - \beta^s (k-m) s_{k,m} + \beta^s(k-m+1) s_{k,m-1} - \gamma^s m s_{k,m} + \gamma^s(m+1) s_{k,m+1}, \label{seqnsApp}
\end{equation}
 where $m$ is in the range $0,\ldots,k$ for each $k$-class in the network (and adopting the convention $s_{k,-1}\equiv s_{k,k+1}\equiv 0)$.
Applying identical arguments, \emph{mutatis mutandis}, to the set $I_{k,m}$, we derive the corresponding system of equations for $i_{k,m}(t)$:
\begin{equation}
\frac{d}{d t}i_{k,m} = -R_{k,m} i_{k,m} + F_{k,m} s_{k,m} - \beta^i (k-m) i_{k,m} + \beta^i(k-m+1) i_{k,m-1} - \gamma^i m i_{k,m} + \gamma^i(m+1) i_{k,m+1}, \label{ieqnsApp}
\end{equation}
for $m=0,\ldots,k$ and for each $k$-class in the network, with time-dependent rates $\beta^i$ and $\gamma^i$ defined though $s_{k,m}$ and $i_{k,m}$ as
\begin{equation}
\beta^i = \frac{\sum_k P_k \sum_{m=0}^k m\, F_{k,m} \,s_{k,m}}{\sum_k P_k \sum_{m=0}^k m\, s_{k,m}}
\quad \text{ and } \quad \gamma^i =  \frac{\sum_k P_k \sum_{m=0}^k m\, R_{k,m} \,i_{k,m}}{\sum_k P_k \sum_{m=0}^k m\, i_{k,m}} .
\label{betai}
\end{equation}

The master equations (\ref{seqnsApp}) and (\ref{ieqnsApp}), with the time-dependent rates $\beta^s$, $\gamma^s$, $\beta^i$ and $\gamma^i$ (defined as nonlinear functions of $s_{k,m}$ and $i_{k,m}$), form a closed system of deterministic equations which, along with initial conditions (\ref{ICs}), can be solved numerically using standard methods \footnote{Mathematica (\texttt{www.wolfram.com}) files for implementing and solving the master equations are available from the author upon request.}. Note that the evolution equations are completely prescribed by the functions $F_{k,m}$ and $R_{k,m}$, and so this method can be applied to any stochastic dynamical process defined by transition rates $F_{k,m}$ and $R_{k,m}$. For the SIS model, equations (\ref{seqnsApp}) and (\ref{ieqnsApp}) were derived in \cite{Marceau10} (see also \cite{Noel09}), with additional terms to study adaptive rewiring of the network.

\section{Appendix B: Homogeneous pair approximation}
The pair approximation (\ref{PA}) derived in the main text is of the type dubbed ``heterogeneous PA'' in \cite{Pugliese09}, because the system includes variables $p_k$ and $q_k$ for each degree class $k$.  A more parsimonious set of equations may be derived under the assumptions of ``homogeneous PA'', wherein the $k$-dependence of edge-based variables $p_k$ and $q_k$ is neglected. As discussed in \cite{Pugliese09}, the reduction in the number of variables typically comes at the cost of reduced accuracy.

For homogeneous PA, the parameter $p(t)$ (resp. $q(t)$) is defined as the probability that  a randomly-chosen neighbor of a susceptible (resp. infected) node is infected. Noting that $p(t)$ can be expressed in terms of $s_{k,m}$ as
\begin{equation}
p(t) = \frac{\sum_k P_k \sum_{m=0}^k m\, s_{k,m}}{\sum_k P_k \sum_{m=0}^k k\, s_{k,m}}=\frac{\sum_k P_k \sum_{m=0}^k m\, s_{k,m}}{\sum_k P_k \, k(1-\rho_k)},
\end{equation}
an evolution equation for $p$ may be derived by multiplying equation (\ref{seqnsApp}) by $P_k\,m$ and summing over $m$ and $k$. The right-hand-side of the resulting equation contains higher moments of $s_{k,m}$, so a closure approximation is needed to proceed. If, similar to the steps yielding equations (\ref{PA}), we here make the ansatz that $s_{k,m}$ and $i_{k,m}$ are proportional to binomial distributions:
\begin{equation}
s_{k,m}\approx (1-\rho_k)\,B_{k,m}(p), \quad\quad i_{k,m} \approx \rho_k\, B_{k,m}(q), \label{ansatz}
\end{equation}
then an equation for $dp/dt$ may be found in terms of only $\rho_k$, $p$, and $q$. Applying the same ansatz to expression (\ref{star}) gives the algebraic relation
\begin{equation}
(1-q)\omega = p(1-\omega) \label{algebraic}
\end{equation}
between $p$ and $q$, where $\omega=\sum_k \frac{k}{z} P_k \,\rho_k$ can be interpreted as the probability that the node at one end of a randomly-chosen edge is infected. After some algebra, we obtain the \emph{homogeneous pair approximation}, consisting of the $k_\text{max}+2$ differential equations:
\begin{eqnarray}
\frac{d}{d t}\rho_k &=& -\rho_k \sum_m R_{k,m} B_{k,m}(q) + (1-\rho_k)\sum_m F_{k,m}B_{k,m}(p) \quad\text{ for }k=0,\ldots,k_{\text{max}}, \nonumber\\
\frac{d}{d t}p &=& \frac{1}{1-\omega}\sum_k \frac{k}{z}P_k\sum_m \left(1+p-2\frac{m}{k}\right)\left((1-\rho_k)F_{k,m}B_{k,m}(p)-\rho_k R_{k,m}B_{k,m}(q)\right), \label{homPA}
\end{eqnarray}
along with the algebraic relation (\ref{algebraic}), and initial conditions $\rho_k(0)=p(0)=\rho(0)$.

 For the SIS model, the homogeneous PA equations (\ref{homPA}) are identical to those recently derived by House and Keeling \cite{House11}, while for the voter model equations (\ref{homPA}) are equivalent to those in \cite{Vazquez08} (in the $N\to\infty$ limit). Note that on $z$-regular graphs, the heterogeneous and homogeneous pair approximations are identical.

\end{document}